\documentclass[12pt]{article}
\usepackage{amsmath,amsfonts,amssymb,latexsym}
\usepackage{epsfig}
\usepackage{graphicx}
\usepackage{wrapfig}
\usepackage{subfigure}
\usepackage{cite}

\newcommand{\sect}[1]{\setcounter{equation}{0}\section{#1}}

\def\ds{\displaystyle}
\def\be{\begin{equation}}
\def\ee{\end{equation}}

\def\bea{\begin{eqnarray}}
\def\eea{\end{eqnarray}}
\def\bean{\begin{eqnarray*}}
\def\eean{\end{eqnarray*}}
\def\N{{\mathbb N}}
\def\Z{{\mathbb Z}}
\def\R{{\mathbb R}}

\def\E{{\mathbb E}}
\def\a{{\alpha}}
\def\b{\beta}

\begin{document}

%%%%%%%%%%%%%%%%%%%%%%%%%%%%%%%%%%%

\thispagestyle{empty}
\hfill \today

\vspace{2.5cm}

\begin{center}
\bf{\LARGE
Lie Groups of 
 Jacobi polynomials \\[0.125cm]
   and \\[0.475cm]
   
   Wigner $d$-matrices 
}
\end{center}

\bigskip\bigskip

\begin{center}
E. Celeghini$^{1,2}$, M.A. del Olmo$^2$ and M.A. Velasco$^2$\footnote{Present address: {\sl CIEMAT, 
Madrid, Spain.}}
\end{center}

\begin{center}
$^1${\sl Dipartimento di Fisica, Universit\`a  di Firenze and
INFN--Sezione di
Firenze \\
I50019 Sesto Fiorentino,  Firenze, Italy}\\
\medskip

$^2${\sl Departamento de F\'{\i}sica Te\'orica and IMUVA, Universidad de
Valladolid, \\ 
E-47011, Valladolid, Spain.}\\
\medskip

{e-mail: celeghini@fi.infn.it, olmo@fta.uva.es}

\end{center}

\bigskip

\bigskip

\begin{abstract}
A symmetry $SU(2,2)$ group in terms of ladder 
 operators is presented for the Jacobi polynomials, $J_{n}^{(\a,\b)}(x)$,
and the  Wigner $d_j$-matrices where the spins $j=n+(\a+\b)/2$ integer and half-integer are considered together.  A unitary irreducible representation of $SU(2,2)$ is constructed and
subgroups of physical interest are discussed.

The Universal Enveloping Algebra of  $su(2,2)$ also allows to construct group structures ($SU(1,1)$, $ SO(3,2)$, $ Spin(3,2)$) whose representations separate integers and half-integers values of 
the spin $j$.

Appropriate $L^2$--functions spaces are realized inside the support spaces of all these representations. Operators acting on these $L^2$--functions spaces belong thus to the  corresponding Universal Enveloping Algebra.

\end{abstract}

\vskip 1cm

\noindent Keywords: Special functions, Jacobi polynomials, Wigner $d_j$-matrices, Lie algebras, Square-integrable functions  

\vfill\eject

%%%%%%%%%%%%%%%%%%%%%%%%%%%%%%%%%%%%%% INTRODUCTION  %%%%%%%
%%%%%%%%%%%%%%%%%%%%%%%%%%%%%%%%%%%%%%%%%%%%%%%%%%%%%%%%%%%%%
\sect{Introduction}\label{intro}

Many attempts have been done to find  a wide but not too inclusive class of functions that can be defined 
``special'', where ``special'' means something more that  
``useful''~\cite{berry2001}.

The actual main  line of work for a possible unified theory of special functions is  the Askey scheme  that
is based on the analytical theory of linear differential equations  \cite{andrews1999,heckman1994,koekoek2010}. 

However, a possible alternative point of view 
was established by employing considerations that belong to a field of mathematics seemingly quite far from them: the theory of representations of Lie groups. This way was  introduced by Wigner~\cite{wigner1955}  and 
Talman~\cite{talman1968} and
later developed mainly by Miller~\cite{miller1968} and 
Vilenkin and Klimyk~\cite{vilenkin1968,vilenkin1991,vilenkin1995}.
In this line, previous papers by us  shown  a direct connection between  some special functions and  well defined Lie groups. The starting point of our work has been the paradigmatic  example of Hermite functions that are a basis on the Hilbert space of the square integrable functions
defined on the configuration space $\R$. As it is well known in the algebraic discussion of the harmonic oscillator,
besides the configuration basis, $\{|x\rangle\}_{x\in \R}$ ,
a discrete basis, $\{|n\rangle\}_{n\in \N}$ -- related to the  Weyl-Heisenberg group $H(1)$ -- can be introduced such that Hermite functions are the transition matrices from one basis to the 
other.
This scheme has been generalized to all the orthogonal polynomials we have, up to now, considered: Legendre and Laguerre polynomials~\cite{celeghini2013a}, associated Legendre polynomials and Spherical Harmonics~\cite{celeghini2013b}. We discuss here the Lie group properties of  Jacobi polynomials and Wigner $d_j$-matrices.

Starting from the seminal work by Truesdell~\cite{truesdell1948}, where   a sub-class of special functions is defined by means of a  set of  formal properties,  we proposed   a possible definition of a fundamental
sub-class of special functions, called ``algebraic special functions" (ASF). These ASF look to be strictly
related to the hypergeometric functions but are  constructed from  the following algebraic assumptions:
\begin{enumerate}
\item 
A set of differential recurrence  relations exists on these ASF that can be associated to a set of ladder operators that span a Lie algebra.
\item 
These ASF support an irreducible representation of this algebra.
\item
A Hilbert space can be constructed on these ASF where the ladder operators have the  hermiticity properties appropriate for constructing a  unitary irreducible representation (UIR) of the associated Lie group.
\item
Second order differential equation that define the ASF can be reconstructed  from all diagonal elements of the Universal Enveloping Algebra (UEA) and, in particular, from
the second order Casimirs of all subalgebras and  of the whole  algebra.
\end{enumerate} 
From these assumptions, we have that:
\begin{itemize}
\item[\; i)]
Applying the exponential map to ASF different sets of functions can be constructed. If the transformation is unitary another algebraically equivalent basis of the Hilbert space is obtained.
When the transformations are not unitary, as in the case of coherent states, sets with different properties are found (like  overcomplete sets).
\item[ii)]
 The ASF are also a basis of an appropriate set of $L^2$--functions   and of an appropriate Hilbert space functions.  This, combined with the above properties, implies that the vector space of the operators operating on $L^2$ (or Hilbert) space functions is homomorphic to the  UEA built on the algebra.
  \end{itemize}

 In \cite{celeghini2013a}  it has been shown that
 Hermite, Laguerre and Legendre polynomials are ASF such that 
 the Hermite  functions  support a UIR of  the Weyl-Heisenberg group $H(1)$ with Casimir ${\cal C} = 0$, while  Laguerre functions and Legendre polynomials are both bases for the 
 UIR of  $SU(1, 1)$ with ${\cal C}=-1/4$. 
Since
Hermite functions are a  basis of square-integrable  functions    defined on the real line, as well as 
Laguerre functions on the semi-line   and  Legendre polynomials on the finite interval~\cite{Cambianis},
all operators acting on such $L^2$--functions can be written inside  the universal enveloping algebra (UEA) of $h(1)$ or $su(1,1)$.
All these properties have been shown to not be restricted to the
above mentioned 1-rank algebras (and groups). Indeed in  \cite{celeghini2013b}
Associated Legendre polynomials and Spherical Harmonics  are shown to share  the same 
properties with  underlying  Lie group $SO(3,2)$ that is of rank 2 like two,  ${l}$ and ${m}$,
are the label parameters of these functions.

Here we present a further confirmation of this scheme  in terms of
the Jacobi polynomials and Wigner $d_j$-matrices showing that they satisfy  the required conditions to be considered ASF and share
the same properties.
Indeed they can be 
associated to well defined ``algebraic Jacobi functions''  (AJF) that
support a UIR of $SU(2,2)$ i.e.,  a Lie group of rank 3 like three are the parameters, $(n, \alpha,\beta)$, of the Jacobi polynomials $J_{n}^{(\a,\b)}(x)$ and three $(j,m,q)$  are also the ones of $d^j_{q,m}$ matrices.

From an applied point of view for both, AJF and Wigner 
$d_j$--matrices, the relevant group chains are  $SU(2,2)\supset SU(2)\otimes SU(2)\supset SU(2)$ to consider together integer and half-integer spin $j$
and  $SU(2,2)\supset SU(1,1)$ to describe them separately.

The paper is organized as follows. 
Section~\ref{algebraicJacobifunctions} is devoted to present the main properties of  the AJF relevant for our discussion and their relations with the  Wigner $d_j$-matrices.
In section~\ref{SUA(2)otimesSUB(2)}   we study the symmetries of   the AJF that keep invariant the principal parameter $j$ and change $m$ and/or $q$. We prove that these ladder operators determine a $su(2)\oplus su(2)$ algebra,
that  allows us to  build up  
a family of UIR of  the group $SU(2)\otimes SU(2)$, i.e.  $U^j\otimes U^j$.
In section~\ref{su11groups} we
construct, by means of four new sets of ladder operators that change the three parameters 
$j, m$ and $q$ in $\pm 1/2$,  each of them generating a  $su(1,1)$ algebra to which infinitely many UIR's  of $SU(1,1)$ 
-- supported by the AJF --  are  associated.   
In Sect.~\ref{su22section} we show that the    ladder operators, obtained in previous sections,  span a $su(2,2)$ Lie algebra and the AJF and Wigner 
$d_j$-matrices generate a UIR of $SU(2,2)$ characterized by the eigenvalue of the quadratic Casimir 
${\cal C}_{SU(2,2)}=-3/2$. Next  Section shows the  AJF and Wigner $d_j$--matrices with labels   all integers and half-integers can be classified in  different and disconnected UIR representations of $SU(1,1)$ or $SO(3,2)$. In section~\ref{operatorsl2} the homomorphism between the space of the operators on the $L^2$ space 
and the  UEA of $su(2,2)$ is discussed. 
Finally  some conclusions and comments  are included.

A previous  unpublished version of this work containing part of the results here presented can be found in \cite{celeghini2013c}.

%%%%%%%%%%%%%%%%%%%%%%%%%%%%%%%%%%%%
%%%%%%%%%%%%%%%%% SECTION 2 %%%%%%%%%%%%
\sect{Algebraic Jacobi functions and their operatorial structure}
\label{algebraicJacobifunctions}

We consider
 the Jacobi polynomials, $J_{n}^{(\a,\b)}(x)$, of degree $n\in \N$  as defined  in terms of the hypergeometric functions\; $_2F_1$ \cite{NIST,luke1969} by
\be\label{j}
J_{n}^{(\a,\b)}(x)=\;\frac{(\a+1)_n}{n!}\;\;\;_2F_1\left[-n,1+\a+\b+n;\a+1;\frac{1-x}{2}\right],
\ee
where  $(a)_n:= a\, (a+1)\cdots (a+n-1)$  is the
 Pochhammer symbol,
or equivalently by \cite{abramowitz1972}
\be\label{jj}
J_{n}^{(\a,\b)}(x)=\;\frac{(\a+1)_n}{n!}\;\left(\frac{1+x}{2}\right)^n\;\;_2F_1\left[-n,-n-\b;\a+1;\frac{x-1}{x+1}\right].
\ee
From \eqref{jj}  an explicit polynomial expression can be obtained \cite{biedenharn1981} 
\be\label{jjj}
J_{n}^{(\a,\b)}(x)=\;
\sum_{s=0}^{n}\;\left(\begin{array}{c} n+\a\\ s\end{array}\right)\;\left(\begin{array}{c} n+\b\\ n-s\end{array}\right)\;
\left(\frac{x+1}{2}\right)^{s}\;\left(\frac{x-1}{2}\right)^{n-s}\;\;.
\ee
where we have considered a generalized binomial coefficient
\[
\left(\begin{array}{c} a\\ s\end{array}\right)
:=\frac{(a+1-s)_s}{s!},
\]
being $a$ an  arbitrary number and $s$ a positive integer.

\begin{figure}
\centerline{\psfig{figure=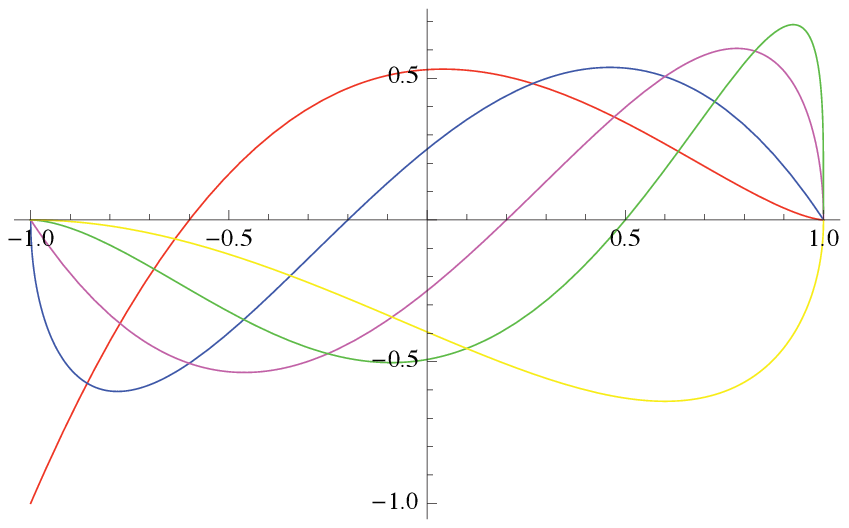,height=8.0cm}}
\caption{\small ${\cal J}_{\frac52}^{\frac32,\frac32}$ (red), ${\cal J}_{\frac52}^{\frac32,\frac12}$ (blue), ${\cal J}_{\frac52}^{\frac32,-\frac{1}{2}}$ (magenta), ${\cal J}_{\frac52}^{\frac32,-\frac32}$ (green), 
${\cal J}_{\frac52}^{\frac32,-\frac52}$ (yellow).} \label{fig_0}
\end{figure}

However to obtain objects  
related to an algebraic structure
like in  \cite{celeghini2013a,celeghini2013b}, we define -alternatively to $n, \a, \b$- three other  (discrete) variables and include an $x$-depending factor. 
We  first substitute
$(n,\a,\b)$ with   $(j, m, q)$ 
\[
 j:=n+\frac{\a+\b}{2}\,, \qquad  m:=\frac{\a+\b}{2}\,,\qquad  q:=\frac{\a-\b}{2}\,,
\]
hence
\[
 n=j - m, \qquad  \a= m+q,\qquad  \b= m-q\; .
\]
Then,  in a second and final step,  we include a $x$-depending factor
related with the integration measure of the Jacobi polynomials. So,
  the fundamental objects of this paper,  we  call
``algebraic Jacobi functions'' (AJF), have the final  form
\be\label{AJF}
{\cal J}_j^{m,q}(x):=
\sqrt{\frac{ \Gamma(j+m+1) \,  \Gamma(j-m+1)}
{ \Gamma(j+q+1) \, \Gamma(j-q+1)}}\,
\left(\frac{1-x}{2}\right)^{\frac{m+q}{2}}\, \left(\frac{1+x}{2}\right)^{\frac{m-q}{2}}
 \,J_{j-m}^{(m+q,m-q)}(x), 
\ee
%%%%%%%%%%%%%%%%%%%%%%
\begin{figure}
\centerline{\psfig{figure=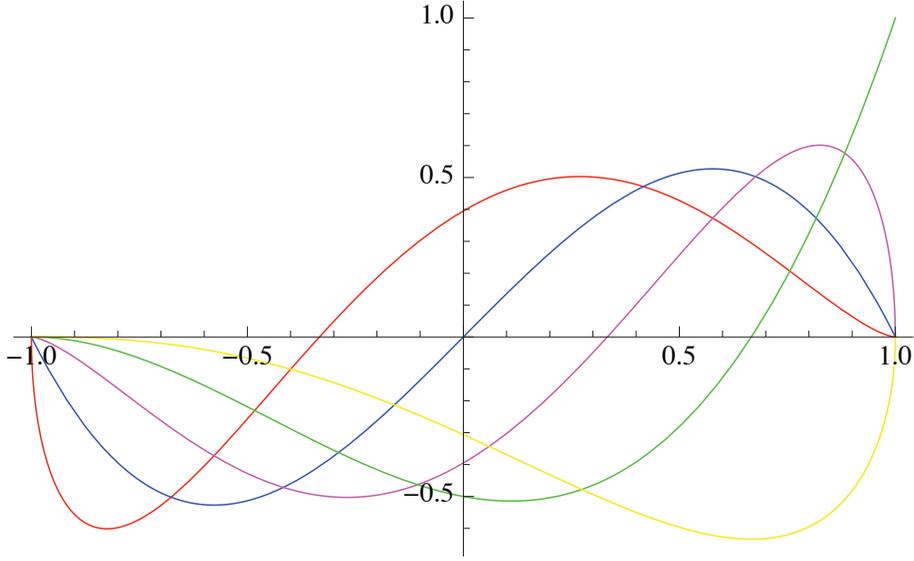,height=8.0cm}}
\caption{\small ${\cal J}_3^{2,1}$ (red), ${\cal J}_3^{2,0}$ (blue), ${\cal J}_3^{2,-1}$ (magenta), ${\cal J}_3^{2,-2}$ (green), ${\cal J}_3^{2,-3}$ (yellow).} \label{fig_00}
\end{figure}
%%%%%%%%%%%%%%%%%%%%%%
\noindent
where in order to obtain  a group representation, as we will prove later,  we have to impose the following restrictions for  $(j,m,q)$
\be\label{conditions}
j\geq |m|, \qquad j\geq |q|, \qquad  2j\in \N,\quad  j-m\in \N,\;\; j-q\in \N
\ee  
Hence, $(j,m,q)$ are all together integers or half-integers 
(see Fig.~\ref{fig_00} and Fig.~\ref{fig_0} where some AFS for different integers and half-integers values of $(j,m,q)$ are displayed).

The conditions \eqref{conditions} rewritten in terms of the original parameters $(n,\a,\b)$ are
\[
n\in \N, \qquad \a,\b \in \Z,\qquad  
\a\geq -n, \qquad
\b\geq -n, \qquad
\a+\b\geq -n.
\]
Note that usually the Jacobi polynomials $J_{n}^{(\a,\b)}(x)$   are defined for
$\a > -1$ and $\b > -1$ ($\a,\,\b \in \R$) in such a way that a unique weight function $w(x)$ allows their
normalization. However (see also 
\cite{biedenharn1981}~p.~49) we have to change such  
restrictions   since we introduce the normalization inside the
 functions and  the algebra requires eq.~(\ref{conditions}). 
In principle
 -because from  the definition \eqref{AJF}  ${\cal J}_j^{m ,q}(x)=0$  for   $j<|q|$-
  the AJF could  be extended to $j<|m|$ considering
\[
\hat{\cal J}_j^{m,q}(x):=\lim_{\varepsilon\to 0} {\cal J}_j^{m+\varepsilon,q}(x) 
\]
as
\[
\hat{\cal J}_j^{m,q}(x)=\left\{
\begin{array}{cl}
  {\cal J}_j^{m ,q}(x)\quad &\forall \,(j,m,q) \;\;\text{verifying all conditions  \eqref{conditions}} \\[0.3cm]
 0 &{j<|m|} 
  \end{array}\right. .
  \]
 
%%%%%%%%%%%%%%%%%%%%%%
%%%%%%%%%%%%%%%%%%%%%%
 
 The AJF  \eqref{AJF} reveal  additional symmetries 
 hidden inside the Jacobi polynomials.
 Indeed we have
\be\label{symmj}
\begin{array}{lll}
{\cal J}_j^{m,q}(x)={\cal J}_j^{q,m}(x),  \\[0.3cm]
{\cal J}_j^{m,q}(x)=(-1)^{l-m}\, {\cal J}_j^{m,-q}(-x),\\[0.3cm]
{\cal J}_j^{m,q}(x)=(-1)^{l-q} \,{\cal J}_j^{-m,q}(-x),  \\[0.3cm]
{\cal J}_j^{m,q}(x)=(-1)^{m+q}\, {\cal J}_j^{-m,-q}(x)\,.
\end{array}
\ee
The proof of these properties is straightforward. The first one can be proved taking into account the following property of the Jacobi polynomials for integer coefficients $(n,\a,\b)$ \cite{biedenharn1981}
\[\label{symmjacobi}
\begin{array}{lll}
{J}_n^{\a,\b}(x)=\ds {\frac{(n+\a)! \,(n+\b)!}{
n! \,(n+\a+\b)!}}\,\left(\frac{x+1}{2}\right)^{-\b}\,{J}_{n+\b}^{\a,-\b}(x) 
\,.
\end{array}
\]
The second property can be derived from the well-known symmetry of the Jacobi polynomials \cite{NIST}
\[
J_{n}^{(\a,\b)}(x)= (-1)^n J_{n}^{(\b,\a)}(-x) ,
\]
and the last two properties   can be proved using the first two  ones.

These AJF   for $m$ and $q$ fixed verify the orthogonality relation
\be 
\label{com}
\ds\int_{-1}^{1}\,{\cal J}_j^{m,q}(x)\;(j+1/2)\; {\cal J}_{j'}^{m,q}(x) \, dx= 
\delta_{j\, j'}
\ee 
as well as
\be\label{completitud}   
\sum_{j={\it sup}(|m|, |q|)}^\infty  {\cal J}_j^{m,q}(x)\; \left(j+1/2\right)\; {\cal J}_j^{m,q}(y) =  \delta(x-y) .
 \ee
Both relations are similar to those of  the Legendre polynomials \cite{celeghini2013a} and  the
associated Legendre polynomials  \cite{celeghini2013b}: all are   orthonormal only up to the factor $j+1/2$. 

The Jacobi equation 
\[
\left[ (1-x^2) \frac{d^2}{dx^2} -((\a+\b+2)x+(\a-\b)) \frac{d}{dx} + n(n+\a+\b+1)\right]\,
J_n^{(\a,\b)}(x)=0 
\]
rewritten in terms of these new functions ${\cal J}_j^{m,q}(x)$ and  of the new parameters $(j,m,q)$ becomes 
\be\label{jacobiequation}
\left[-(1-x^2)\,\frac{d^2}{dx^2}+2\,x\frac{d}{dx}+ 
\frac{2\;m\;q\;x+m^2+q^2}{1-x^2}- j(j+1)\right] \,{\cal J}_j^{m,q}(x)=0\; ,
\ee
where the symmetry under the interchange $m \Leftrightarrow q$ is evident. 

It is worthy noticing  that the AJFs \eqref{AJF}, with the substitution $x = \cos\b$,  are essentially the  elements of the Wigner $d_j$ rotation 
matrices~\cite{wigner27,biedenharn1981} 
\[\label{dj}
d^j_{q m}(\b)=\sqrt{\frac{(j+m)!\,(j-m)!}{(j+q)!\,(j-q)!}}\,(\sin\b/2)^{m-q}\,(\cos\b/2)^{m+q}\,J^{(m-q,m+q)}_{j-m}(\cos\b)
\]
that verify the conditions \eqref{conditions}.
The explicit relation between them is
\be\label{relation}
d^j_{q m}(\b)= {\cal J}^{m,-q}_{j}(\cos\b).
\ee
Eqs.~\eqref{symmj} are equivalent to the well-known relations among the $d^j_{qm}$ like, for instance, 
\[
d^j_{q m}(\b)=(-1)^{q-m}\,d^j_{m q}(\b).
\] 
The paper could be rewritten in terms of the $d^j_{q m}$ but the formal complexity should be larger and, perhaps, this is the reason why the lying algebras were not known up to now.

Coming back to the AJF,  the starting point for   finding  the algebra  is now
the construction of the rising/lowering differential operators that change the labels   $(j,m,q)$ of the  AJF
 by 1 or 1/2. The fundamental limitation of this analytical approach is  that the indices 
are considered as parameters \cite{miller1968} 
 that,  in iterated applications, must be introduced by hand.
This problem has been  solved in  \cite{celeghini2013a} 
where a consistent vector space framework 
was introduced
to allow the iterated use of recurrence formulas by means of operators of which the parameters involved are eigenvalues.

Indeed -- in order to display the  operator structure on the set 
$\{{\cal J}_j^{m,q}(x)\}$ --  we introduce,
in consistency with the quantum theory approach, not only the 
operators $X$ and $D_x$ of the configuration space, such that 
\[
X\, f(x) = x\, f(x) ,\qquad D_x\, f(x) = f'(x),\qquad [X,D_x]= -1\, ,
\]
but  also three other operators $J$, $M$ and $Q$ such that 
\be\label{jmq}
J\; {\cal J}_j^{m,q}(x) = j\; {\cal J}_j^{m,q}(x) \quad\; M\, {\cal J}_j^{m,q}(x) = \, m\; {\cal J}_j^{m,q}(x)\,
\quad\; Q\, {\cal J}_j^{m,q}(x) = \, q\; {\cal J}_j^{m,q}(x),  
\ee
i.e. diagonal on the algebraic Jacobi functions and, thus, 
belonging -in the whole algebraic scheme- to the
Cartan subalgebra.

%%%%%%%%%%%%%%%%%%%%%%%%%%%%%%%%%%%%%%%%%%%
%%%%%%%%%%%%%%%%%%%%%%%SECTION 3 %%%%%%%%%%%%

\sect{Algebras  for Jacobi functions with $\Delta j = 0$}\label{SUA(2)otimesSUB(2)}

We start from the differential-difference 
equations and the difference equations verified by the Jacobi functions (a  complete list of which can be found in
Refs.~\cite{NIST,luke1969,abramowitz1972}).
The  procedure is  laborious, so that,
we only sketch the simplest case with $\Delta j = 0$, related with 
$su(2)$ and well-known for the $d_j$  in terms of the angle \cite{wu}.
  
  Let us start from the operators that change the values of $m$ only.
Consider the equations (18.9.15) and  (18.9.16)  of Ref.~\cite{NIST}
\[\begin{array}{rll}
\ds \frac{d}{dx} J_{n}^{(\alpha , \beta)}(x) &=&\ds \frac{1}{2}(n+\alpha+\beta+1)\,
J_{n-1}^{(\alpha+1,\beta+1)}(x)\;,
\\[0.3cm]
\ds\frac{d}{dx}\left[(1-x)^\alpha (1+x)^\beta J_{n}^{(\alpha , \beta)}(x)\right] &=&\ds
 -2(n+1) (1-x)^{\alpha-1} (1+x)^{\beta-1}\,J_{n+1}^{(\alpha-1, \beta-1)}(x),
\end{array}\] 
which are far to be symmetric, 
but   allow us to define the operators 
\be\label{Adiferential} 
A_{\pm} :=\; \pm \,\sqrt{1-X^2}\, D_x\,
+\, \frac{1}{\sqrt{1-X^2}}\; (X M+Q),
\ee
that act on the  algebraic Jacobi functions ${\cal J}_j^{m,q}(x)$   as  
\be\label{actionA}
A_\pm\; {\cal J}_j^{m,q}(x) = \sqrt{(j\mp m)\,(j \pm m+1})\;
\,{\cal J}_{j}^{m\pm 1,\, q}(x) .
\ee

The operators (\ref{Adiferential})  are a generalization for $Q\neq 0$ of the
operators $J_\pm$ introduced  in Ref.~\cite{celeghini2013b}  
for the associated Legendre functions related  
to the
AJF  with $q=0$.  Moreover 
eqs.~(\ref{actionA}), that are independent from $q$,
coincide with eqs.~(2.11) and  (2.12) of Ref.~\cite{celeghini2013b}.  

Defining  now $A_3:= M$ and taking into account the action of the operators $A_\pm$ and $A_3$ on the Jacobi functions, eqs. 
\eqref{actionA} and \eqref{jmq}, it is easy to check that 
$A_\pm$ and $A_3$ close a $su(2)$ algebra, that commutes with  $J$ and $Q$,  denoted in the following by $su_A(2)$
\[\label{suA2}
[A_3,A_\pm]=\pm A_\pm \qquad [A_+,A_-]=2 A_3 . 
\]

Thus,  the  algebraic Jacobi functions $\{{\cal J}_j^{m,q}(x)\}$ such that  $2 j \in \N$,
$j-m \in \N$ and  $-j\leq m\leq j  $  support  the  $(2 j+1)$-dimensional UIR of the
Lie group  $SU_A(2)$ independently from the value of $q$.

Similarly to  \cite{celeghini2013b}, starting from 
the differential realization  \eqref{Adiferential}
of the $A_\pm$ operators 
 we recover the Jacobi differential equation \eqref{jacobiequation}  from the 
 Casimir, ${\cal C}_A$, of $su_A(2)$ 
\be\label{eqcasimirA}
 \left[{\cal C}_A-J(J+1)\right]\;{\cal J}_j^{m,q}(x)\equiv 
 \left[A_3^2+\frac12\{A_+,A_-\}-J(J+1)\right]\,{\cal J}_j^{m,q}(x) = 0\;.
 \ee
 Indeed, eq.~\eqref{eqcasimirA} reproduces the operatorial form of \eqref{jacobiequation}, i.e.
\be\label{jacobiequationoperator}
\left[-(1-X^2) D^2_x + 2 X D_x + \frac{1}{1-X^2} (2 X M Q +M^2 + Q^2) - J(J+1)\right] {\cal J}_j^{m,q}(x) = 0 .
\ee
On the other hand  we can make use of
the factorization method \cite{schrodinger,infeld-hull1951},   relating  second order differential equations to  product   of first order ladder operators in such a way that  the application of the first operator modifies the values of the parameters of the second one. 
Taking into account this fact,  iterated application of   
\eqref{Adiferential} gives the two equations
 \be \begin{array}{l}\label{jacobinodiferential1}
\left[A_+\,A_- -(J+M)\,(J-M+1)\right]\; {\cal J}_j^{m,q}(x)=0\;,
\\[0.3cm]
\left[A_-\,A_+ -(J-M)\,(J+M+1)\right]\; {\cal J}_j^{m,q}(x)=0\;,
\end{array}\ee
that reproduce again the  operator form \eqref{jacobiequationoperator} of the Jacobi equation.
Equations \eqref{eqcasimirA}, \eqref{jacobinodiferential1}  are particular cases of a general rule:  the defining Jacobi equation can be recovered appliying to 
${\cal J}_j^{m,q}$ the Casimir operator   of any  involved algebra and sub-algebra  as well as  any diagonal product of ladder operators.

Now, using the symmetry $m \Leftrightarrow q$ of the ${\cal J}_j^{m,q}(x)$, we construct
the algebra that changes $q$ leaving $j$ and $m$ unchanged. From   $A_\pm$ two  new operators $B_\pm$ are thus  defined
\be \label{Bdiferential}
B_{\pm} := \pm \,\sqrt{1-X^2}\, D_x+\frac{1}{\sqrt{1-X^2}} \,   (X Q+M)\,,
\ee
and their action on the Jacobi functions  is
\be \label{Bnodiferential}
B_\pm\,{\cal J}_j^{m,n}(x) = \sqrt{(j\mp q)\,(j \pm q+1})\; 
\,{\cal J}_{l}^{m,q\pm 1}(x) .
\ee
Obviously also the operators $B_\pm$ and $B_3:=Q$\, close a $su(2)$ algebra   we denote $su_B(2)$
\[
[B_3,B_\pm]=\pm B_\pm \qquad [B_+,B_-]=2 B_3 ,
\]
and the Jacobi functions $\{{\cal J}_j^{m,q}(x)\}$  with $2j\in \N$, $j-q\in \N$ and $-j\leq q \leq j$ close the  $(2 l+1)$-dimensional UIR of the Lie group  $SU(2)_B$ independently from the value of $m$.

Again we can recover the Jacobi equation  \eqref{jacobiequationoperator}
from the Casimir, ${\cal C}_B$, of  $su_B(2)$ 
\[
\left[{\cal C}_B-J(J+1)\right] {\cal J}_j^{m,q}(x) = 
\left[  B_3^2 +      \frac12\{B_+,B_-\}-J(J+1)\right] {\cal J}_j^{m,q}(x)=0
\]
as well as from the diagonal second order products 
\[ \begin{array}{l}
\left[B_+\,B_- -(J+Q)\,(J-Q+1)\right]\; {\cal J}_j^{m,q}(x)=0\;,
\\[0.3cm]\left[B_-\,B_+ -(J-Q)\,(J+Q+1)\right]\; {\cal J}_j^{m,q}(x)=0\;.
\end{array}\]

\begin{figure}
\centerline{\psfig{figure=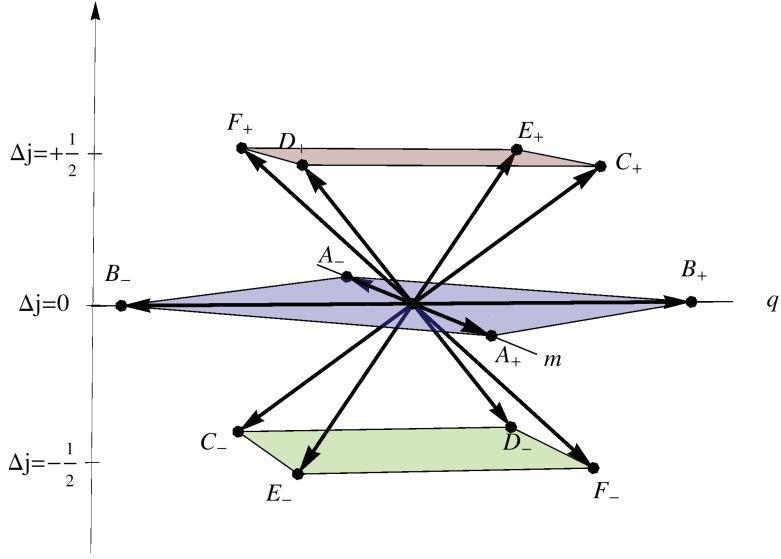,height=8.0cm}}
\caption{\small Action of the ladder operators  on the Jacobi functions ${\cal J}_j^{(m,q)}(x)$ represented by the triplets $(j,m,q)$. The planes displayed correspond to the pairs $(m,q)$, the parameter $j$ (or $\Delta j$) is represented in the vertical axis.} \label{fig_1}
\end{figure}

A more complex algebraic scheme  appears 
in common applications of the operators $A_\pm$ and $B_\pm$.  As the operators $\{A_\pm, A_3\}$  commute with $\{B_\pm,B_3\}$,  the algebraic structure is the  direct sum of 
the two Lie algebras
\[
su_A(2)\oplus su_B(2) .
\]
A new symmetry of the  AJFs emerges in the space of ${\cal J}_j^{m,q}(x)$ with   $j$ fixed. For $(j, m, q)$, all integer or half-integer, formulae 
\eqref{actionA}, \eqref{Bnodiferential} and \eqref{jmq}  are the well known expressions for the infinitesimal generators
 of the group $SU_A(2)\otimes SU_B(2)$.
 The Jacobi functions 
${\cal J}_j^{m,q}(x)$ for fixed $j$ and $-j \leq m\leq j$, $-j \leq q\leq j$ determine a UIR  of this group. 
From \eqref{Adiferential} and \eqref{Bdiferential},
taking into account that always the operators $M$ and $Q$  have been written at the right
of $X$ and $D_x$, it can be shown that $A^\dagger_\pm=A_\mp,\;B^\dagger_\pm=B_\mp$
and the representation  is, as required,  unitary. In Fig.~\ref{fig_1}   the action of the operators $A_\pm,B_\pm$ on the parameters $(j,m,q)$ that label the Jacobi functions  corresponds to the plane $\Delta j=0$.

%%%%%%%%%%%%%%%%%%%%%%SECTION 4 %%%%%%%%%%%%%%%%%%
%%%%%%%%%%%%%%%%%%%%%%%%%%%%%%%%%%%%%%%%%%%%%%%
\sect{Other ladder operators inside algebraic Jacobi functions and $su(1,1)$ representations}\label{su11groups}

We mentioned before that  many difference and differential-difference relations for Jacobi polynomials are known~\cite{luke1969, NIST}.  
Starting from them a su(2,2) Lie algebra can be constructed.  It 
has fifteen infinitesimal generators, where three of them are Cartan generators (for instance, $J, M$ and $Q$). As the four generators   that commute with 
$J$ ($A_\pm$   and $B_\pm$) have been introduced in the preceeding paragraph, we have to construct eight non-diagonal
operators more. They are:
\be\label{Differential8}
\displaystyle
\begin{array}{l}
C_+:=\,+\dfrac{(1+X)\sqrt{1-X}}{\sqrt{2}}\, D_x - \dfrac{1}{\sqrt{2(1-X)}} \left(X\,(J+1)-(J+1+M+Q)\right),
\\[0.5cm]
C_-:=-\,
\dfrac{(1+X)\sqrt{1-X}}{\sqrt{2}}\, D_x - \dfrac{1}{\sqrt{2(1-X)}}\; \left(X\, J- (J+M+Q)\right),
\\[0.5cm]
D_+:=-\,
\dfrac{(1-X)\sqrt{1+X}}{\sqrt{2}}\, D_x + \dfrac{1}{\sqrt{2\,(1+X)}}\; \left(X (J+1)+ (J+1+M-Q)\right),
\\[0.5cm]
D_-:=+\,
\dfrac{(1-X)\sqrt{1+X}}{\sqrt{2}} \, D_x + \dfrac{1}{\sqrt{2\,(1+X)}}\; \left(X\,J + (J+M-Q)\right),
\\[0.5cm]
E_+:=-\,
\dfrac{(1-X)\sqrt{1+X}}{\sqrt{2}}\, D_x+
\dfrac{1}{\sqrt{2\,(1+X)}}\left( X (J+1)+(J+1-M+Q) \right),
\\[0.5cm]
E_-:=+\,\dfrac{(1-X)\sqrt{1+X}}{\sqrt{2}}\, D_x+\dfrac{1}{\sqrt{2(1+X)}}\; \left( X J+(J-M+Q) \right),
\\[0.5cm]
F_+:=-\,
\dfrac{(1+X)\sqrt{1-X}}{\sqrt{2}}\, D_x+
\dfrac{1}{\sqrt{2\,(1-X)}}\left( X (J+1)-(J+1-M-Q) \right),
\\[0.5cm]
F_-:=+\,
\dfrac{(1+X)\sqrt{1-X}}{\sqrt{2}}\, D_x+\dfrac{1}{\sqrt{2\,(1-X)}}\,
 \left( X J-(J-M-Q) \right).
\end{array}
\ee
%%%%%%%%%%%%%%%%%%%%%%
%%%%%%%%%%%%%%%%%%%%%%

All these differential operators act on the space $\{{\cal J}_j^{m,q}\}$ for $j,m,q$ integer and half-integer such that $j\ge |m|,|q|$. The explicit form of their action is: 
\be\label{Nodifferential8}
\displaystyle
\begin{array}{rcl}
C_+\,{\cal J}_j^{m,q}(x)&=&\sqrt{(j+m+1)(j+q+1)}\; {\cal J}_{j+1/2}^{m+1/2,\;q+1/2}(x),\\[0.4cm]
C_-\,{\cal J}_j^{m,q}(x)&=&\sqrt{(j+m)(j+q)}\; {\cal J}_{j-1/2}^{m-1/2,\;q-1/2}(x),\\[0.4cm]
D_+\,{\cal J}_j^{m,q}(x)&=&\sqrt{(l+m+1)(l-q+1)}\; {\cal J}_{j+1/2}^{m+1/2,\;q-1/2}(x)\\\\
D_-\,{\cal J}_j^{m,q}(x)&=&\sqrt{(j+m)(j-q)}\; {\cal J}_{j-1/2}^{m-1/2,\,q+1/2}(x,)\\[0.4cm]
E_+\,{\cal J}_j^{m,q}(x)&=&\sqrt{(j-m+1)\,(j+q+1)}
\;{\cal J}_{j+1/2}^{m-1/2,\;q+1/2}(x),\\[0.4cm]
E_-\,{\cal J}_j^{m,q}(x)&=&\sqrt{(j-m)\,(j+q)}\;
{\cal J}_{j-1/2}^{m+1/2,\;q-1/2}(x),\\[0.4cm]
F_+\,{\cal J}_j^{m,q}(x)&=&\sqrt{(j-m+1)\,(j-q+1)}\;
\,{\cal J}_{j+1/2}^{m-1/2,\,q-1/2}(x),\\[0.4cm]
F_-\,{\cal J}_j^{m,q}(x)&=&\sqrt{(j-m)\,(j-q)}\;
{\cal J}_{j-1/2}^{m+1/2,\;q+1/2}(x).
\end{array}
\ee
From \eqref{Nodifferential8} (or \eqref{Differential8} remembering the order in the operators) we have
\[
C_\pm^\dagger = C_\mp,\qquad 
 D_\pm^\dagger = D_\mp,\qquad
  E_\pm^\dagger = E_\mp,\qquad
   F_\pm^\dagger = F_\mp,
\]
i.e. all these rising/lowering  operators have the hermiticity properties required by the representation to be unitary.
The operators   \eqref{Differential8}  change all parameters by $\pm 1/2$.  We define all together $X_\pm$ 
the ones that change $j$ in $j\pm1/2$, that in Fig.~\ref{fig_1}     correspond to the planes $\Delta j=\pm1/2$. 

From the eqs.~\eqref{Differential8}   it is easily stated that
\be\begin{array}{llr}\label{weylsym}
 D_\pm (X,D_x,M,Q)  &=  &C_\pm (-X,-D_x,M,-Q),\\[0.3cm]
E_\pm (X,D_x,M,Q)  &=  &C_\pm (-X,-D_x,-M,Q),\\[0.3cm]
F_\pm (X,D_x,M,Q)  &=  &- C_\pm (X,D_x,-M,-Q) .
\end{array}\ee
Thus, because of the Weyl symmetry of the roots,  we limit ourselves   to discuss the operators 
$C_\pm$. 
Taking thus into account  their action  on the Jacobi functions 
  we get
\be\label{suC11}
 [C_+,C_-]= -2 C_3,  \qquad   [C_3,C_\pm]=\pm C_\pm 
\ee
where 
\be \label{c3spectrum}
C_3:= J+ \frac{1}{2}(M+Q)+\frac12.
\ee 
Hence $\{C_\pm,C_3\}$ close a $su(1,1)$ algebra that is denoted  $su_C(1,1)$.

As in the  cases of the operators $A_\pm$ and $B_\pm$, we obtain the Jacobi differential equation  up to a non vanishing factor
  from the Casimir ${\cal C}_C$ of $su_C(1,1)$,  written in terms of
  \eqref{Differential8} and \eqref{c3spectrum},
  \be\label{casimirsuC11}
  {\cal C}_C{\cal J}_j^{m,q}(x)\equiv\left[C_3^2-\frac12\{C_+,C_-\}\right]{\cal J}_j^{m,q}(x)=\frac{1}{4}\left[(m+q)^2-1\right]{\cal J}_j^{m,q}(x).
  \ee
 Indeed 
\be\begin{array}{l}\label{eqcasimirC}
 \ds\left[{\cal C}_C-\frac{1}{4}(M+Q)^2+\frac{1}{4}\right]\;{\cal J}_j^{m,q}(x)\\[0.5cm]
\qquad \equiv 
 \left[C_3^2-\frac12\{C_+,C_-\}-\frac{1}{4}\,(M+Q)^2+1/4\right]\,{\cal J}_j^{m,q}(x) =0  
\end{array}\ee
allows us to  recover the Jacobi eq.~\eqref{jacobiequationoperator}.
Analogously  the same result derives from eqs.
\be\begin{array}{l}\label{Cjacobinodiferential1}
\left[C_+\,C_- -(J+M)\,(J+Q)\right]\; {\cal J}_j^{m,q}(x) =0 ,
 \\[0.4cm]
\left[C_-\,C_+ -(J+1+M)\,(J+1+Q)\right]\; {\cal J}_j^{m,q}(x)=0 , 
 \end{array}\ee
obtained by the factorization method.

From \eqref{eqcasimirC} we see that since $(m+q) = 0,\pm 1,\pm 2,\pm 3,\cdots $ the IR of su(1, 1) with 
${\cal C}_C =(m+q)^2/4-1/4 = -1/4, 0, 3/4,2, 15/4, \cdots$ are obtained. 
Hence,  the set of AJF 
  supports many
 infinite dimensional UIR, of $SU(1, 1)$ of the discrete series for $SU_C(1,1)$~\cite{Bargmann47}.

Similar results can be found for the other  ladder operators $D\pm,
E\pm,F\pm$ with the substitutions  \eqref{weylsym} in  all  eqs.~(\ref{suC11}-\ref{Cjacobinodiferential1}).

%%%%%%%%%%%%%%%%%%% SECTION 5 %%%%%%%%%%
%%%%%%%%%%%%%%%%%%%%%%%%%%%%%%%%%%%
\sect{The complete symmetry group of  $\{{\cal J}_j^{m,q}(x)\}$: $SU(2,2)$}\label{su22section}

If one represents the action of the twelve operators $A_\pm , B_\pm , C_\pm , D_\pm , E_\pm ,
F_\pm$, that we have defined in previous sections, we obtain Fig.~\ref{fig_1}.
To obtain the root system of the simple Lie algebra $A_3 \equiv D_3$ we have only simply to add three points in the origin
corresponding to the elements  $J, M$ and  $Q$ of the Cartan subalgebra.

The Lie commutators of the generators $A_\pm,B_\pm,C_\pm,D_\pm,E_\pm,F_\pm,J,M,Q$ are 
\[\begin{array}{llll}
[J,A_\pm]=0,\quad &
[J,M]=0,\quad &[J,B_\pm]=0,\quad &[J,Q]=0,
\\[0.4cm]
[J,C_\pm]=\pm \frac12 \,C_\pm,\quad &
[J,D_\pm]=\pm \frac12 \,D_\pm,\quad &[J,E_\pm]=\pm \frac12 \,E_\pm,\quad &[J,F_\pm]=\pm \frac12 \,F_\pm,
\\[0.4cm]
  [M,B_\pm]=0,\quad & [M,Q]=0,&&
\\[0.4cm]
[M,C_\pm]=\pm\frac12\, C_\pm,\quad &[M,D_\pm]=\pm\frac12\, D_\pm,\quad &[M,E_\pm]=\mp\frac12\, E_\pm,\quad &[M,F_\pm]=\mp\frac12\, F_\pm,
\\[0.4cm]
  [Q,A_\pm]=0,\quad &&&
\\[0.4cm]
[Q,C_\pm]=\pm \frac12 \,C_\pm,\quad &[Q,D_\pm]=\mp \frac12 \,D_\pm,\quad &[Q,E_\pm]=\pm \frac12 \,E_\pm,\quad &[Q,F_\pm]=\mp \frac12 \,F_\pm,
\\[0.4cm]
[A_+,A_-]=2 A_3,\quad &[A_3,A_\pm]=\pm A_\pm,  &(A_3=M) , & 
\\[0.4cm]
[B_+,B_-]=2 B_3,\quad &[B_3,B_\pm]=\pm B_\pm, &(B_3=Q), &
\\[0.4cm]
[C_+,C_-]=-2 C_3,\quad & [C_3,C_\pm]=\pm C_\pm, &(C_3= J+ \frac{1}{2}(M+Q)+\frac12) ,&
\\[0.4cm]
[D_+,D_-]=-2 D_3,\quad &[D_3,D_\pm]=\pm D_\pm, &(D_3= J+ \frac{1}{2}(M-Q)+\frac12),&
\\[0.4cm]
[E_+,E_-]=-2 E_3,\quad & [E_3,E_\pm]=\pm E_\pm, &(E_3= J+ \frac{1}{2}(-M+Q)+\frac12),&
\\[0.4cm]
[F_+,F_-]=-2 F_3,\quad & [F_3,F_\pm]=\pm F_\pm, &(F_3= J-\frac{1}{2}(M+Q)+\frac12),&
\\[0.4cm]
[A_\pm,B_\pm]=0,\quad &[A_\pm,B_\mp]=0,\quad & &
\\[0.4cm]
[A_\pm,C_\pm]=0,\quad &[A_\pm,C_\mp]=\pm E_\mp,\quad &[A_\pm,D_\pm]=0,\quad & [A_\pm,D_\mp]=\mp F_\mp,
\\[0.4cm]
[A_\pm,E_\pm]=\pm C_\pm,\quad &[A_\pm,E_\mp]=0,\quad &[A_\pm,F_\pm]=D_\pm,
\quad &[A_\pm,F_\mp]=0,
\\[0.4cm]
[B_\pm,C_\pm]=0,\quad &[B_\pm,C_\mp]=\mp D_\mp,\quad &[ B_\pm,D_\pm]=\pm C_\pm,
\quad &[B_\pm,D_\mp]=0,
\\[0.4cm]
[B_\pm,E_\pm]=0,\quad &[B_\pm,E_\mp]=\mp F_\mp ,\quad &[B_\pm,F_\pm]=\pm E_\pm,
\quad &[B_\pm,F_\mp]=0,
\\[0.4cm]
[C_\pm,D_\pm]=0,
\quad &[C_\pm,D_\mp]=\mp B_\pm,\quad
&[C_\pm,E_\pm]=0,
\quad &[C_\pm,E_\mp]=\mp A_\pm,
\\[0.4cm]
[C_\pm,F_\pm]=0,
\quad &[C_\pm,F_\mp]=0, &&
%\\[0.4cm]
\end{array}\]
\[\begin{array}{llll}
[D_\pm,E_\pm]=0,
\quad &[D_\pm,E_\mp]=0,\quad
&[D_\pm,F_\pm]=0,
\quad &[D_\pm,F_\mp]=\mp A_\pm,
\\[0.4cm]
 [E_\pm,F_\pm]=0,\quad & [E_\pm,F_\mp]=\mp B_\pm . &&
\end{array}\]

The quadratic  Casimir of $su(2,2)$ has the form 
\[\begin{array}{ll}
{\cal C}_{su(2,2)}&=\ds
\frac12\left(\{A_+,A_-\}+\{B_+,~B_-\}-\{C_+,C_-\}-\{D_+,D_-\}  -\{E_+,E_-\}-\{F_+,F_-\}\right)\\[0.3cm]
&\qquad\qquad\ds+\,
\frac{1}{2}\,\left( A_3^2 + B_3^2 + C_3^2 + D_3^2 + E_3^2+ F_3^2 \right)\\[0.4cm]
&=\ds
\frac12\left(\{A_+,A_-\}+\{B_+,~B_-\}-\{C_+,C_-\}-\{D_+,D_-\}  -\{E_+,E_-\}-\{F_+,F_-\}\right)\\[0.3cm]
&\qquad\qquad\ds+\,2 J(J+1)+M^2+Q^2+\frac12\\[0.4cm]
&\ds \equiv -\frac{3}{2} .
\end{array}\]
From it and taking into account the differential realization  of the operators involved,
\eqref{Adiferential},  \eqref{Bdiferential} and \eqref{Differential8},  we recover again the Jacobi equation  
\eqref{jacobiequationoperator}.

\begin{figure}
\centerline{\psfig{figure=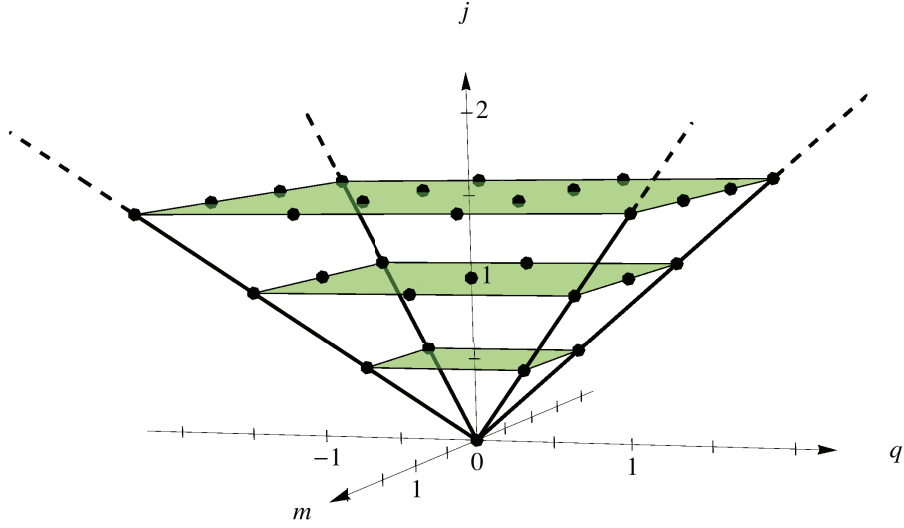,height=8.0cm}}
\caption{\small IR of $su(2,2)$ supported by the AJF ${\cal J}_l^{m,q}(x)$ represented by the black points. The horizontal planes correspond to IR of $su_A(2)\oplus su_B(2)$.} \label{fig_3}
\end{figure}

Hence, the AJF support a UIR of the group $SU(2,2)$ with the value -3/2 of  
${\cal C}_{su(2,2)}$ (see Fig.~\ref{fig_3}). Also, as we have seen along the previous sections, the Jacobi equations is recovered form the Casimir of any subalgebra of  ${su(2,2)}$ as well as from any diagonal product of ladder operators. 

In this UIR  of $SU(2,2)$ the integer and half-integer values of $(j, m, q)$ are putted all together.  The symmetries where integer and half-integer values belong to
different UIR  will be discussed in next Section. 

%%%%%%%%%%%%%%%%%%%%%%
%%%%%%%%%%%%%%%%%%%%%%

 %%%%%%%%%%%%%%%%%%%%%%%%%%%%%%%%%%%%%%%%%%%%%%%%%%%%%%%%%%%%%%%%%%%%%%%%%%

\sect{Symmetries inside integers/half-integers labels}
\label{symmetriessu11}
  
 In the preceding sections we have obtained a set of symmetries on the AJF 
 ${\cal J}_j^{m,q}(x)$ where the operators $A_\pm$ ($B_\pm$)   change only the  label $m$ ($q$) of $\pm 1$  leaving invariant the remaining two labels and the other operators $X_\pm$ of the algebra  change all the $(j,m,q)$
by a half-integer quantity. Integer and half-integer values of $(j,m,q)$ are indeed related to a unique UIR of $SU(2,2)$. 

Now, composing the action ot pairs of operators $X_\pm$\,, we can 
construct two operators that change $(j,m,q)$ to $(j\pm1, m ,q)$.
Indeed $F_\pm C_\pm$ (equivalently, $C_\pm F_\pm$,  $D_\pm E_\pm$ or
$E_\pm D_\pm$)  are such  operators. Moreover they,  in general second
order differential operators, reduce --because of the Jacobi equation  (\ref{jacobiequationoperator})-- to a first order
differential operators when applied to the ${\cal J}_j^{m,q}$.

As discussed before\, the ${\cal J}_j^{m,q}$ have $j\geq|m|$ and $j\geq|q|$, but now we have to consider separately 
 specific values of $m$ and $q$.

To begin with let us start with  states such that  $j\geq|m| > |q|$ .
From (\ref{Differential8}) and (\ref{jacobiequationoperator})
we get that the two conjugate hermitian operators
 \be\label{Kpm}
K_+:=   F_+ C_+  \frac{1}{\sqrt{(J+1)^2-Q^2}} ,
\qquad
K_-:= F_- C_- \frac{1}{\sqrt{J^2-Q^2}} 
\ee
 can be written on the AJF as
 \bea\label{nJmasmenosoperators}
 K_+ &:=&\left(-(1-X^2) \;D_x+  X\;( J+1)+
 \frac{MQ}{J+1}\;\right)   \frac{J+1}{\sqrt{(J+1)^2-Q^2}} \\[0.3cm]
K_- &:=& \left((1-X^2) \;D_x+  X\; J+\frac{MQ}{J}\right) \frac{J}{\sqrt{J^2-Q^2}} \;\label{K-}
\eea
always well defined because $j>|q|$ on these AJF.   By inspection, 
their action does not depend from the value of $q$:
 \be\begin{array}{lll} \label{actionK}
 K_+\;  {\cal J}_j^{m,q}(x) &=&\, \sqrt{(j+1)^2-m^2}\; 
 {\cal J}_{j+1}^{m,q}(x) ,
\\[0.3cm]
\ds K_-\; {\cal J}_j^{m,q}(x) &=&\, \sqrt{j^2-m^2}\;  {\cal J}_{j-1}^{m,q}(x)  ,
\end{array}\ee
and $K_\pm$ together with $K_3:=J+1/2$ close a 
$su(1,1)$ Lie algebra
\[
[K_+,K_-]=-2 K_3,\qquad [K_3,K_\pm]=\pm K_\pm
\]
  such that the ${\cal J}_j^{m,q}$ with $m$ and $q$ fix and\, $j\geq |m| > |q|$\, are a basis of the UIR of $su(1,1)$  with Casimir
\[
{\cal C} = m^2 -1/4 .
\]

For the states where\, $|m|<|q|$\, the procedure is analogous:
we have only to interchange $M \Leftrightarrow Q$
 in   $K_\pm$. 

The problem is more complex when $|m|=|q|$ as the action of $K_-$ 
is not well defined in eq.(\ref{K-})
for $j=|m|=|q|$. 
To extend its definition to this case we have to consider not the
eigenvalues of  $M$ and $Q$ but their limit\, i.e. 
\[\ds
K_- := \lim_{\epsilon \to 0} 
\left[ 
\left((1-X^2) \,D_x+  X\, J+
\frac{(M+\epsilon)(Q+\epsilon)}{J}
\right)
\frac{J}{\sqrt{J^2-(Q+\epsilon)^2}}\; \right].
\]
In this way the action of $K_-$ does not change for $j > |m|= |q|$ while  for $j=|m|=|q|$,
results from the product of a first factor that goes like 
$\epsilon$ an a second one that goes like $\epsilon^{-1/2}$  
and it is thus zero.  

In conclusion, all\, ${\cal J}_j^{m,q}$\, close, for fixed $m$ and $q$, 
a UIR of $SU(1,1)$.
If $|m|\geq |q|$  we have eqs.~(\ref{actionK})
with $j=|m|, |m|+1, |m|+2 \dots$  and  Casimir invariant 
${\cal C} = m^2-1/4$
while, for $|m| < |q|$, we have to exchange everywhere $q$ and  $m$.

Note that, unlike the UIR of $SU(2,2)$, each representation of $SU(1,1)$ contains only states with integer or
half-integer values of its labels.
This is particularly relevant for the $d^j_{q\, m}$ as usual
symmetries in physics do not mix integer and half-integer spins\, i.e. bosons and fermions.

This $SU(1,1)$ cannot in general be extended to larger groups as 
its operators cannot be combined with other operators 
to construct a bigger algebra, but  there are few exceptions.
Indeed when $q=0$, i.e. on the ${\cal J}_j^{m, 0}$ with $j, m \in \Z$ and $j\geq |m|$,
we can define not only the $\{K_\pm, K_3\}$ but also the $\{A_\pm, A_3\}$ and thus
the whole algebra $so(3,2)$ described in \cite{celeghini2013b}.
The $\{{\cal J}_j^{m,0}\}$ are indeed related to the
Associate Legendre Polynomials $P_{n+\a}^{\a}$
and thus a basis of a UIR of $SO(3,2)$.
By inspection $K_\pm$ are exactly the ones reported in fomulas (2.13-14) of \cite{celeghini2013b} and $A_\pm$ are the
$J_\pm$ of formulas (2.11-12) of the same paper. 
Because of the symmetry $m \Leftrightarrow q$ we have a $SO(3,2)$ also
for $m=0$ i.e. on the $\,\{{\cal J}_j^{0,q}\}\,$.

Also "fermions"  states
$\{{\cal J}_j^{m,\pm1/2}\}$ and $\{{\cal J}_j^{\pm1/2,q}\}\,$
are related to the same algebra $so(3,2)$ but, of course, to a representation of its covering group
$Spin(3,2)$ \cite{spin32}.

%%%%%%%%%%%%%%%%%%%%%%%%%%%%%%%%%%%%%%%%%%%%%%%%%%%%%%%%%%%%%%%

 \sect{$L^2$--functions spaces and ${\cal J}_j^{m,q}(x)$}
\label{operatorsl2}

As already discussed in details in \cite{celeghini2013a} and \cite{celeghini2013b}, special functions together with
their Lie group properties play a role in $L^2$--functions and Hilbert spaces. 

Formulas (\ref{com}) and (\ref{completitud}) where $m$ and $q$ are fixed allow to introduce $j$ and $x$ as conjugate variables on
the same Hilbert space.
Let us thus begin discussing the case of section \ref{symmetriessu11},
where $m$ and $q$ are fixed parameters  characterizing the UIR of $su(1,1)$ and its
support  Hilbert space.

Operators  (\ref{Kpm}) written  as
(\ref{actionK})   on the $\{{\cal J}_j^{m,q}(x)\}$
allow to  define, in abstract form, for fixed $m$ and $q$,
the action on a Hilbert space constructed in the space of
eigenvectors\, $\{|j, (m,q)\rangle\}$\, of the operators $J$\,
as
 \be\begin{array}{lll} \label{actionKH}
 K_+\;  |j, (m,q)\rangle &=&\, \sqrt{(j+1)^2-m^2}\; 
 |j+1, (m,q)\rangle ,
\\[0.3cm]
\ds K_-\; |j, (m,q)\rangle &=&\, \sqrt{j^2-m^2}\;  |j-1, (m,q) \rangle  .
\end{array}\ee
and $\{|j, (m,q)\rangle\}$ is a basis of the Hilbert space\, i.e.
\be\label{comH}
 \langle j, (m,q)| j', (m,q) \rangle  = \delta_{j\, j'} \qquad
\sum_{j={\it sup}(|m|, |q|)}^\infty  | j, (m,q) \rangle\; \langle j, (m,q)| = {\cal I} .  
\ee
This allows to define
\be\label{bb}
|x, (m,q)\rangle := \;
\sum_{j={\it sup}(|m|, |q|)}^\infty  
 |j, (m,q)\rangle\, \sqrt{j+1/2}\;\, {\cal J}_j^{m,q}(x)
\ee
where the vectors  $\{|x, (m,q)\rangle\}$ are a basis of
the configuration space  $\E=(-1,1)\subset\R$  
i.e.   
\[
\langle x, (m, q)| x', (m, q)\rangle = \delta(x-x')\quad \qquad 
\int_{-1}^{+1} |x, (m, q) \rangle\, dx\, \langle x, (m, q)|\; =\, {\cal I}\;.
\]  
This implies that  the $\{{\cal J}_j^{m,q}(x)\}$ are the transition matrices between the two bases: 
\[
{\cal J}_j^{m,q}(x)\, =\,\frac{1}{\sqrt{j+1/2}}\;  \langle x, (m,q)|j, (m,q)\rangle\, =\, 
\frac{1}{\sqrt{j+1/2}}\;\langle j, (m,q)|x, (m,q)\rangle \;. 
\]
and 
\be\label{123}
|j,(m,q)\rangle = \int_{-1}^{+1} |x,(m,q)\rangle \,\sqrt{j+1/2}\; \,{\cal J}_j^{m,q}(x) \,dx\;.
\ee

As $m$ and $q$ are invariant of the representation,
this situation is similar  to the Lagrange functions discussed in  \cite{celeghini2013a}, where only  two conjugate variables
$j$ and $x$ (one discrete and one continuous) are present.
The difference is that in  \cite{celeghini2013a}  a basis was associated to each
element $\mathfrak g$ of the group $SU(1,1)$,  while here we have a basis for each element of the 
triple $(m, q, \mathfrak{g})$.

The case of two discrete variables, related to the group $SO(3,2)$, has been discussed in \cite{celeghini2013b}
and it will be not reconsidered here. On the contrary we consider  the new case of three variables 
case, related to $SU(2,2)$ where both $m$ and $q$ are modified by the algebra.

The Hilbert space is now
 $\E\times \Z \times \Z/2$ , where $\E$ is again $(-1,1)\subset\R$, 
$
\Z/2 := \{ 0, \pm 1/2, \pm1, \pm3/2, \pm 2, \cdots\}
$
is related to $m$ and $ \Z $ to $m-q$, as $m$ and $q$ are together 
integer or half-integer.
The space $\E\times \Z \times \Z/2$,  with  basis  $\{|x, m,q\rangle\}$, is the  direct sum of 
the Hilbert spaces $\E_{m,q}$ with $m$ and $q$  fixed,
\[
\E\times\Z\times\Z/2 = \bigcup_{m-q\in\Z} \; \bigcup_{q\in\Z/2}  \;\E_{m,q}.
\]
  
Orthonormality and 
completeness are now,
\[
\langle x,m,q | x', m',q' \rangle\; =\; \delta(x-x')\, \delta_{m\, m'}\,\, \delta_{q\,q'} ,\qquad
\sum_{m,q }\, \int_{-1}^{+1}  |x,m,q\rangle\, dx\, \langle x,m,q|\; =\; {\cal I}.
\]

Analogously to (\ref{123})  
we can now define inside the Hilbert space a new basis  $\{|j,m,q\rangle\}$ with $m,q\in\Z/2,\; j \geq |m|, \; j \geq |q|,\;j-m\in\N,\;j-q\in\N$ by
\[
|j,m,q\rangle := \int_{-1}^{+1} |x,m,q\rangle \,\sqrt{j+1/2}\; \,{\cal J}_j^{m,q}(x) \,dx\;.
\]
so that
\[
\langle j,m, q' | j', m', q' \rangle\; =\; \delta_{j\,j'}\, \delta_{m\, m'}\, \delta_{q,\, q'},
\qquad
\sum_{ j, m, q}  |j,m, q\rangle\;   \langle j,m, q|\; =\; {\cal I} .
\]
The $\{{\cal J}_j^{m,q}(x)\}$ plays  the role of transition matrices: 
\[
{\cal J}_j^{m,q}(x) =\frac{1}{\sqrt{j+1/2}}\;  \langle x,m,q|j,m,q\rangle = 
\frac{1}{\sqrt{j+1/2}}\;\langle j,m,q|x,m,q\rangle \;. 
\]

Like in Ref.~\cite{celeghini2013a, celeghini2013b}
the role of the algebraic Jacobi functions $\{{\cal J}_j^{m,q}(x)\}$ as transition matrices reflects the fact that the generators can be seen as differential operators in $\E\times \Z \times \Z/2$,  
or algebraic operators in the spaces of labels   $\N/2\times \N \times \N$, where $\N/2$ is related to $j$
and the two $\N$ to $j-m$ and $j-q$.  This allows to make explicit the Lie algebra structure
in contrast with Ref.~\cite{miller1968,vilenkin1968,vilenkin1991}.

An arbitrary vector 
$|f\rangle \, \in L^2(\E,\Z,\Z/2)$
can be alternatively expressed as
\[
|f\rangle \;=\; \sum_{m,q=-\infty}^{\infty} \int_{-1}^{+1} dx\; |x,m,q\rangle \, f^{m,q}(x) \;=
\sum_{m,q=-\infty}^{+\infty} \;\;\;\sum_{j=sup(|m|,|q|)}^\infty |j,m,q\rangle \;f^{m,q}_j
\]
where 
\[\begin{array}{lllll}
f^{m,q}(x)&:=& \langle x,m,q|f\rangle&= &\ds \sum_{j=sup(|m|,|q|)}^\infty \sqrt{j+1/2}\; {\cal J}_j^{m,q}(x)\, f_j^{m,q} 
,\\[0.5cm]
f_j^{m,q}& :=& \langle j,m,q|f\rangle &=&\ds \int_{-1}^{+1}dx\; 
\sqrt{j+1/2}\;{\cal J}_j^{m,q}(x)\, f^{m,q}(x)\; 
\end{array}\]
and all the $L^2$--functions defined on $(\E,\Z,\Z/2)$ can be written as
\be\label{fl2}   
\sum_{m,q=-\infty}^{\infty}\;\sum_{j=sup(|m|,|q|)}^\infty  \;  \sqrt{j+1/2}\; {\cal J}_j^{m,q}(x)\; f_j^{m,q}  .
\ee

As the\, $\{{\cal J}_j^{m,q}\}$\, are a basis of a UIR of $SU(2,2)$ and\, (see eq.(\ref{fl2})),
at the same time, a basis of the $L^2$--functions defined on 
$(\E,\Z,\Z/2)$ the $L^2(\E,\Z,\Z/2)$
belong to the same UIR of $SU(2,2)$. This implies that every change of basis in the $L^2(\E,\Z,\Z/2)$
is related to an element $\mathfrak g$  of the group $SU(2,2)$ and that every operators that acts on the
$L^2(\E,\Z,\Z/2)$ can be written inside the Universal Enveloping Algebra of $su(2,2)$.

%%%%%%%%%%%%%%%%%%%%%%%%
%%%%%%%%%%%%%%%%%%%%%%
\sect{Conclusions}

The algebraic Jacobi functions are a particular case of the algebraic special functions recently  introduced by us 
\cite{celeghini2013a,celeghini2013b} whose relevance  seems to be related to the following points:
\begin{enumerate}
\item
The role of intertwining between second order differential equations and
Lie algebras played by the algebraic special functions.

\item
Peculiar properties of the algebraic special functions, that perhaps could be assumed as their definition, look to be
 that -- taking into account the fundamental second order differential equation -- all diagonal elements of the UEA can be found to be related to it and that all non-diagonal elements can be written as first order differential operators.
 
\item 
 The fact that ASF (here AJF) are at the same time an 
irreducible representation of a Lie algebra (here $su(2,2)$) and a basis of $L^2$--functions (here the ones defined on 
$\E \times \Z \times \Z/2)$
 allows to establish a homorphism between the UEA of the Lie algebra
and the vector space of the operators defined on the $L^2$--functions. In this way the framework of  Quantum Mechanics is reproduced with 
 $L^2(\E,\Z,\Z/2)$  Hilbert space and  $su(2,2)$-UEA as space of operators acting on it.
 
\item 
As the ASF are a basis of a unitary irreducible representation of the corresponding Lie group also,
all sets obtained from ASF applying an element $\mathfrak g$ of the Lie group
are bases in the space of the $L^2$--functions. Moreover in the 
$SU(1,1)$ cases presented in Sect.~\ref{symmetriessu11} each basis depends not only on the group element $\mathfrak g$ but also on the pair $(m,q)$.

\item
Although the UIR of $SU(2,2)$ put  together integer and half-integer values of the labels $(j,m,q)$, 
the analysis of Sect.~\ref{symmetriessu11} separates boson and fermions systems as usual in physical applications. In this case 
$L^2(\E)$ is the  Hilbert space and  $su(1,1)$-UEA is the  space of operators acting on it.

\item 
The algebraic Jacobi functions have the same symmetry of  the elements of the 
$d_j$-matrices. This enhances the interest of the algebraic special functions  in physical applications since the Wigner matrices play an important role in the description of Quantum Mechanics. 
\end{enumerate}

%%%%%%%%%%%%%%%%%%%%%%%%%
%%%%%%%%%%%%%%% ACKNOWWLEDGMENTS %%%%%%%%%%%%%%%%%%%%%%%%%%%%%%%%

\section*{Acknowledgments}

This work was partially supported  by the Ministerio de
Educaci\'on y Ciencia  of Spain (Projects FIS2009-09002 with EU-FEDER support),  by the
Universidad de Valladolid and by
INFN-MICINN (Italy-Spain).

%%%%%%%%%%%%%%%%%%%%%% BIBLIOGRAPHY %%%%%%%%%%%%%%%%%%%%%%%%%%%%%%

%%%%%%%%%%%%%%%%%%%%%%%%%%%%%%%%%%%
%%%%%%%%%%%%%%%%%%%%%%%%%%%%%%%%%%%


\begin{thebibliography}{99}



\bibitem{berry2001}
M. Berry,  {\it Phys. Today} {\bf  54} (2001) 11.

\bibitem{andrews1999}
G.E.  Andrews, R. Askey, R. Roy, {\it Special Functions}, Cambrige Univ. Press,  Cambridge, 1999.

\bibitem{heckman1994}
G. Heckman, H. Schlichtkrull, {\it Harmonic Analysis and Special Functions on Symmetric Spaces}, Academic Press,  New York,  1994.

\bibitem{koekoek2010}
R. Koekoek,   P.A. Lesky,    R.F. Swarttouw, {\it  Hypergeometric
Orthogonal Polynomials and Their $q$-Analogues}, Springer,  Berlin, 2010  (and references therein).

\bibitem{wigner1955} 
E.P. Wigner, {\it The application of  group theory to the special functions of mathematical physics},  in: Princeton Lectures, 1955.

\bibitem{talman1968} 
J.D. Talman, {\it Special functions: a group theoretic approach}, Benjamin,  New York, 1968.

\bibitem{miller1968}
W.  Miller, {\it Lie Theory and Special Functions}, Academic Press,  New York,  1968.

\bibitem{vilenkin1968}  
N. Ja. Vilenkin, {\it Special Functions and the Theory of Group Representations}, Amer. Math. Soc.,   Providence, 1968.

\bibitem{vilenkin1991}  
N. Ja. Vilenkin,  A.U. Klimyk, {\it Representation of Lie Groups and Special Functions} vols. 1, 2 and 3, Kluwer,  Dordrecht,  1991, 1993 and 1992  (and references therein).

\bibitem{vilenkin1995}  
N. Ja. Vilenkin, A.U. Klimyk, {\it Representation of Lie Groups and Special Functions: Recent Advances}, Kluwer,  Dordrecht, 1995.

\bibitem{celeghini2013a}
 E. Celeghini, M.A. del Olmo,    {\it Ann. of Phys.}  {\bf 335}  (2013)   78.

\bibitem{celeghini2013b}
 E. Celeghini, M.A. del Olmo, {\it Ann. of Phys.}
 {\bf 333}   (2013)    90. 

\bibitem{truesdell1948}  
C. Truesdell, {\it Annals in Math. Studies} {\bf 18}, Princeton Univ. Press,   Princeton,  1948.

\bibitem{Cambianis} 
S. Cambianis, {\it Proc. of the Am. Math. Soc.} {\bf 29}  (1971)  284.

\bibitem{celeghini2013c}
E. Celeghini, M.A. del Olmo, M. A. Velasco, {\it Jacobi Polynomials and $SU(2,2)$},  arXiv:1307.7380.

\bibitem{NIST}
F.W.J. Olver, D.W.  Lozier, R.F.  Boisvert,  C.W.  Clark, {\it NIST Handbook of Mathematical Functions},  Cambridge Univ. Press, New York, 2010. 

\bibitem{luke1969}
Y.L. Luke, {\it The Special Functions and Their Approximations} Vol.1,
 Academic Press, San Diego, 1969 (pp. 275--276).

\bibitem{abramowitz1972}
 M. Abramowitz, I. Stegun, {\it Handbook of Mathematical Functions with Formulas, Graphs, and Mathematical Tables}, Dover,  San Diego, 1972.

\bibitem{biedenharn1981}
L.C. Biedenharn and J.D. Louck, {\sl Angular Momentum in Quantum Mechanics}, Addison-Wesley, Reading, 1981

\bibitem{wigner27}
E. Wigner,  
{\it Zeitschrift f\"ur Physik} {\bf 43} (1927)  624.

\bibitem{wu}  Wu-Ki Tung,  {\it Group Theory in Physics}, World Scientific, Singapore   (1985).

\bibitem{schrodinger}
 E.  Schr\"odinger, {\it Proc. Roy. Irish Acad}. {\bf A46},  183 (1940); {\bf A47} (1941)  53. 

\bibitem{infeld-hull1951}  L. Infeld, T.E. Hull, {\it Rev. Mod. Phys}. 
{\bf 23} (1951)  21. 

\bibitem{Bargmann47} V. Bargmann, {\it Ann. of Math.} {\bf 48} (1947)  368.

\bibitem{spin32}  H.B. Lawson, M.L. Michelsohn, {\it Spin Geometry}, Princeton Univ. Press, Princeton, 1989.


\end{thebibliography}
\end{document}